# Superconducting MgB$_2$ thin film's nano-bridges for cryo-electronic application

M. Gregor [1,2*], R. Micunek [1], T. Plecenik [1], T. Roch [1], A. Lugstein [3], E. Bertagnolli [3], I. Vávra [4], M. Štefecka [2], M. Kubinec [1], M. Leporis [2], V. Gašparík [1,2], P. Kúš [1], and A. Plecenik [1],

[1] Department of Experimental Physics, FMPI CU, Mlynská dolina, 842 48 Bratislava, Slovak Republic,

[2] Biont a.s., Karloveská 63, 842 29 Bratislava, Slovak Republic,

[3] Institute for Solid State Electronics, TU-Wien, Florag. 7, A-1040 Vienna, Austria

[4] Institute of Electrical Engineering, Slovak Academy of Sciences, Dúbravská cesta 9, Bratislava, Slovak Republic

___________________________________________________________________

*Corresponding author: Maroš Gregor, Karloveská 63, 842 29 Bratislava, Slovakia

Tel:+421-2-206 70 744, Fax: +421-2-206 70 748

E-mail: maros.gregor@gmail.com

**Abstract**

Precursor $MgB_2$ thin films were prepared on sapphire substrates by magnetron sputtering. Influence of ex-situ annealing process on superconducting $MgB_2$ thin films roughness is discussed. Optimized annealing process of MgB precursor thin films in vacuum results in smooth superconducting $MgB_2$ thin films with roughness below 10 nm, critical temperature $T_{con}$ = 31 K and transition width ?$T_c$ less than 1 K. Nano-bridges based on the superconducting $MgB_2$ thin films using optical and Focused Ion Beam lithography were prepared. Critical current density $j_c$ (4.2 K) measured on 50 nm wide strip was $7.3 \times 10^6$ A/cm$^2$ and no significant loss of superconducting properties was detected. Resistance vs. temperature and critical current vs. temperature characteristics were measured on these structures using standard DC four probe measurements.

**Introduction**

Cryoelectronic applications of $MgB_2$ superconducting thin films still promise multilateral availing. Good stability, high critical temperature and low anisotropy of this material are advantages in comparison with conventional superconductors. Since 2001 when superconducting properties of $MgB_2$ were discovered, many techniques to prepare thin films were used. Promising results were obtained by molecular beam epitaxy [1], by co-deposition of boron and magnesium from two resistive sources [2], pulsed laser deposition [3], magnetron sputtering [4] and sequential deposition of boron and magnesium. The best results were obtained by pulsed laser deposition [5].

Extensive efforts to realize Josephson junction fabrication technology have been made worldwide. Several excellent reviews on the problems of the Josephson junction fabrication have been published, including planar-type [6] and edge-type [7]. In spite of tunnel-junction domination throughout the history of the Josephson-effect studies, appreciable attention has been attracted to the non-tunnel-type junctions like weak links with variable thickness and width [8]. There have been a number of reports on the fabrication of various types of $MgB_2$ weak links, including point-contact or break junctions [9], nano-bridges, sandwich-type tunnel junctions [10], planar junctions by localized ion damage in thin films [11] and ramp-type junctions [12]. In most cases the roughness of superconducting $MgB_2$ thin films and depression of superconducting properties in tens nanometer scale delimit their application for preparation of sub-micro and nano-structures.

In this paper we focus our attention on preparation of MgB$_2$ thin films with roughness below 10 nm and production of nano-bridges by optical lithography and Focused Ion Beam (FIB) technique. The main accentuation was put on the influence of dimensionality decreasing on the superconducting properties of nano-bridges. The properties of MgB$_2$ thin films and nano-bridges were inspected by Atomic Force Microscopy (AFM), Transition Electron Microscopy (TEM) and DC four probe measurements of resistance vs. temperature (R-T) and critical current vs. temperature (I$_c$-T) characteristics. Preparation of smooth superconducting MgB$_2$ thin films and nano-bridges with non-depressed superconducting properties suitable for preparation of e.g. Superconducting Quantum Interferometer (SQUID), Superconducting Single Photon Detectors (SSPD), etc. are presented.

**Experimental**

Superconducting MgB$_2$ thin films were prepared by magnetron sputtering and ex-situ annealing in argon atmosphere and in vacuum. Before MgB$_2$ thin film preparation, the sapphire substrates were chemically cleaned in acetone, isopropylalcohol and distilled water, and were inserted into vacuum chamber on a rotating holder. Then deposition process was realized from two circular magnetrons with diameter 5 cm. The boron magnetron was fed by RF power supply and the magnesium magnetron was fed by DC power supply. The vacuum chamber was evacuated to overall pressure $5 \times 10^{-6}$ Pa and consecutively filled with argon to working pressure $7.4 \times 10^{-1}$ Pa. Boron was deposited at the power 250 W and magnesium was deposited at the power 20 W. The whole deposition time was 40 minutes and the thickness of MgB precursor thin films was 300 nm.

Two ex-situ annealing processes were realized. The first one was an ex-situ annealing in tubing vacuum furnace. MgB precursor samples were inserted to quartz tube, and the quartz tube was evacuated to pressure 200 Pa and scavenged by argon (Ar) several times. After this procedure the tube was filled with Ar to atmospheric pressure and inserted into furnace. Optimization of annealing process was done in temperature range from 670 °C to 690 °C and time range from 1.5 minutes to 2.5 minutes [2].

The second ex-situ annealing process was performed in high-vacuum chamber at pressure $1 \times 10^{-6}$ Pa. MgB precursor thin films were inserted to annealing vacuum chamber with Boraletric heater. Optimization of the annealing temperature was done at temperature range of 700 – 900 °C and annealing time between 2-4 min. In order to prepare well defined nanostructures, it is necessary to obtain very low surface roughness of prepared thin films. From this point of view, post-deposition annealing process seems to be the most critical one. Thus, an optimizing process was done to determine the smallest roughness and the highest critical temperature $T_c$.

Atomic Force Microscope (NT-MDT Solver P47) has been used for surface roughness analysis. AFM scanning was operated in semi-contact mode using conventional Si tip with high reflectivity Au coating on back side of cantilever for better reflection of laser control system. The maximum lateral scan range of the piezo-tube is 50 × 50 µm$^2$, the maximum vertical range is 2 µm. Typical curvature radius of the tip is 10 nm, and the resonant frequency is 255 kHz.

On MgB$_2$ thin films prepared using in vacuum annealing process, 5 and 10 µm wide strips we prepared by optical litography. For determination of the microstructures the Ar$^+$ ion etching with energy of ions 600 eV and ion current density about 130 µA/cm$^2$ generated by ion gun PLATAR Klan 53M was used. For FIB machining, the MgB$_2$ micro-structructures were coated with 20 nm Au film in vacuum chamber, and on such samples the FIB lithography was applied.

FIB experiments were carried out using the Micrion twin lens FIB system (model 2500) equipped with a Ga liquid metal ion source. The system was operated at an acceleration voltage of 50 kV with selectable 25 µm-beam-limiting aperture corresponding to a beam current of 20 pA. For patterning, the ion beam was raster scanned in a digital scan mode over a defined milling box. The scanning strategy was encoded by the choice of the distance between the pixels, as well as the time each pixel is exposed to the ion beam.

After FIB milling additional Ar$^+$ ion beam etching was applied, in order to remove the Au coating from the strip. On these structures, resistance vs. temperature (R-T) and critical current vs. temperature ($I_c$-T) characteristics were measured using standard DC four probe measurements in transport He Dewar container.

The structure of MgB$_2$ films was investigated by TEM (JEOL 200FX) on cross-sectional specimens prepared by both side ion (Ar) beam (5keV) milling in the Gattan (PIPS691) equipment. The microstructure of the films was inspected at accelerating voltage 200 kV.

**Results and discussion**

MgB precursor thin films with thickness 300 nm were prepared by co-deposition from two magnetrons. Prepared thin films exhibit smooth surface with good adhesion. The ex-situ annealing process of MgB precursor thin films at temperature 680 °C for 1.5 min in Ar atmosphere results in superconducting $MgB_2$ thin films with $T_{con}$ about 33.7 K and zero resistance critical temperature $T_{c0}$ about 32 K (Fig. 1a). The other way of ex-situ annealing process at temperature 850 °C for 3 min in vacuum results in superconducting $MgB_2$ thin films with $T_{con}$ about 31.3 K and zero resistance critical temperature $T_{c0}$ about 30.8 K (Fig. 1b). Differences in transition temperatures to superconducting state are not substantial, however, the main difference is in the quality of the surface.

$MgB_2$ thin films annealed in argon atmosphere exhibit grainy surface with peak-to-peak value of z coordinate up to 260 nm (Fig. 2a, Tab. 1). Even if critical temperature $T_{c0}$ is high, such $MgB_2$ thin films are not suitable for preparation of nanostructures because of their high roughness.

$MgB_2$ thin films annealed in vacuum exhibit three times lower roughness than that ones annealed in argon (Fig. 2b). For additional decreasing of roughness of such thin films additional annealing in $Ar^+$ ion beam was applied. More than twice additional decreasing of the roughness was achieved.

For comparison of the surface roughness of the $MgB_2$ thin films annealed in Ar atmosphere and in vacuum, two different methods were used. The first one describes maximal peak-to-peak position of z coordinates (Tab. 1). Because of existence of random failure on the surface, the peak-to-peak method does not describe adequately the real surface roughness. For this reason, a second statistical processing of the image surface was applied (using "GRAIN ANALYSIS" software included in AFM menu). From experimental data the value of the standard deviation for z-coordinate (Root Mean Square) on the sample surface within the area was calculated as:

$$R_q = \sqrt{\frac{1}{N_x N_y} \sum_{i=1}^{N_x} \sum_{j=1}^{N_y} (z(i,j) - z_{mean})^2}$$

Results of the surfaces annealed in Ar atmosphere, in vacuum and in vacuum with additional $Ar^+$ etching during 30 min are shown in Table 1.

As one can see, conventional annealing in argon atmosphere is highly unsuitable for sub-micrometer scale manufacturing because $R_q$ is about 29.3 nm. After annealing in high vacuum $R_q$ significantly decreases on the value lower than 8 nm, and lower than 1.5 nm after additional $Ar^+$ ion etching (Tab. 1).

For a complex description of the thin film microstructure the TEM analysis was performed in cross-sectional TEM specimens. $MgB_2$ thin films prepared by both methods exhibit nanocrystalline $MgB_2$ phase embedded in amorphous phase. The $MgB_2$ crystallite size is varied from several nanometers to several tens of nanometers. The 011 preferential orientation of $MgB_2$ in the normal to the film plane direction is observed. Electron diffraction exhibits also lines indicating existence of randomly oriented MgO inside the film (Fig. 3). However in $MgB_2$ thin films annealed in vacuum, the MgO-diffraction spots are sufficiently weaker and exhibit homogenous crystalline structure with the size of crystallites up to ten nanometers. The cross-sectional TEM micrographs of $MgB_2$ thin film annealed in vacuum are shown on Fig.4a,b,c. Because no significant differences of superconducting parameters ($T_{con}$ and $?T_c$) between both types of $MgB_2$ thin films were observed, the second type was chosen for preparation of nanostructures, because of its lower roughness.

20-µm-wide strips were prepared by optical lithography on $MgB_2$ thin films. Critical current density $j_c$(4.2 K) of these strips was higher than $5 \times 10^6$ A/cm$^2$. On these microstrips the FIB machining was applied. FIB cuts comprise two 10 µm long and 0.3 µm wide trenches (Fig. 5). The rectangular boxes were achieved by scanning of the focused ion beam in a multiscan raster mode with a pixel spacing of 10 nm, dwell time of 5 µs, and an ion dose of 0.8 nC/µm$^2$.

The spacing between the trenches and the overlap were varied, in order to achieve nano-bridges ranging from 50x50 nm up to 150x150 nm. The patterns were generated outside the field of view to avoid any unintentional ion exposure by imaging. All experiments were carried out under ambient conditions in a clean room environment.

R-T, V-I and $I_c$-T characteristics of nano-bridges were measured using standard DC four probe measurements in transport He Dewar container. Figure 6a shows R-T dependence of 50 nm $MgB_2$ bridge. No crucial changes of critical temperature of $MgB_2$ were observed before and after machining by FIB. Critical current of 50 nm bridge was 0.73 mA (Fig. 6b) and a critical current density $j_c$ (4.2 K) was $7.3 \times 10^6$ A/cm$^2$. This value of critical current density is a little higher than in microbridge. One can explain this phenomenon by decreasing of parallel and series weak link connections between nanocrystallites inside the nano-bridge.

From above mentioned results we can conclude that the surface roughness of $MgB_2$ thin films strongly depends on annealing conditions. $MgB_2$ thin films annealed in vacuum exhibit roughness below 10 nm and superconducting properties of these thin films were not significantly decreased in the bridge of 50 nm scale.

**Conclusion**

The great difference in the surface roughness was observed when using two different ways of annealing. $MgB_2$ thin films annealed in argon atmosphere exhibit surface roughness higher than 100 nm and they are unsuitable for sub-micrometer scale manufacturing. Annealing of MgB thin films in vacuum results in superconducting $MgB_2$ thin films with surface roughness below 10 nm and critical temperature $T_c \sim 31$ K, ? $T_c < 1$ K and critical current density $j_c(4.2\ K) \approx 5\times10^6$ A/cm$^2$. On such type of thin films, the testing nano-bridges were fabricated by Focus Ion Beam. Critical current density $j_c(4.2\ K)$ measured on 50 nm wide strip was $7.3\times10^6$ A/cm$^2$, critical current $T_{con} = 25$ K and ?Tc = 1.5 K.

**Acknowledgments**

This work was supported by the APVT projects No. APVT-51-016604, APVT-20-011804 and project aAV/112 6/2004.

**Figure captions**

Fig.1 R(T) dependences measured on $MgB_2$ thin films a) annealed in argon atmosphere ($T_{con}$ = 33.7 K, $T_{c0}$ = 32 K and $\Delta T_c$ = 1.7 K) and b) annealed in vacuum ($T_{con}$ = 31.3 K, $T_{c0}$ = 30.8 K and $\Delta T_c$ = 0.5 K).

Fig.2 AFM image of thin $MgB_2$ films annealed a) in Ar atmosphere b) in vacuum and c) in vacuum with additional Ar etching for 30 min.

Fig.3 Electron diffraction of $MgB_2$ thin film annealed in vaccum chamber.

Fig.4 The cross-sectional TEM micrographs $MgB_2$ thin film annealed in vacuum.
  a) bright field image shows low surface roughness.
  b) $100_{MgB2}$ and $200_{MgO}$ (overlapping reflex) dark field image reveals the $MgB_2$ and MgO crystallites distribution in the film
  c) $101_{MgB2}$ dark field image reveals the $MgB_2$ crystallites form and their distribution in film.

Fig.5 AFM image of the 50 nm $MgB_2$ nanobridge prepared by FIB milling.

Fig.6  a) Resistance vs. temperature and b) critical current vs. temperature dependences of the 50 nm wide strip prepared by FIB ( $T_{con}$ = 25 K, $\Delta T_c$ = 1.5 K).

**Table captions**

Table1 Comparison of the surface quality of $MgB_2$ annealed in Ar atmosphere and in vacuum.

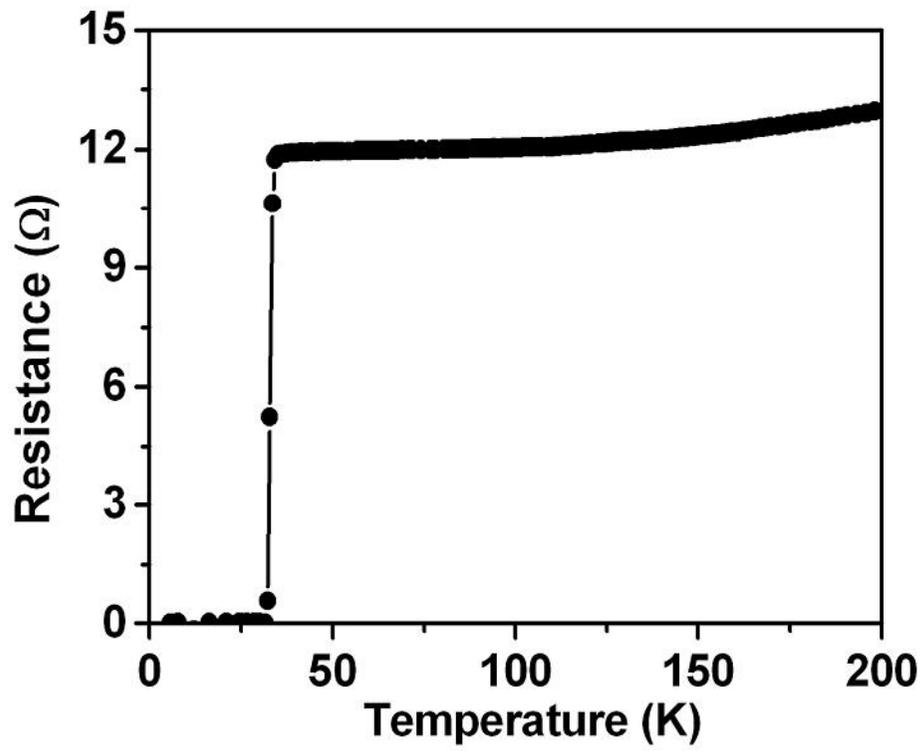

Fig.1a

Gregor et. al

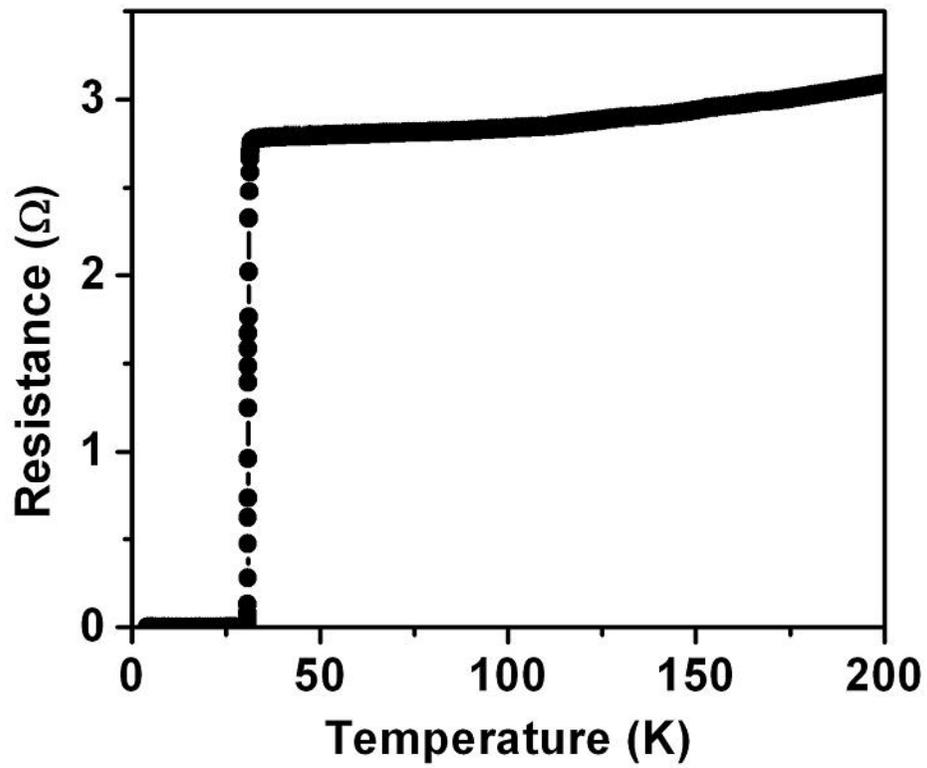

Fig.1b

Gregor et. al

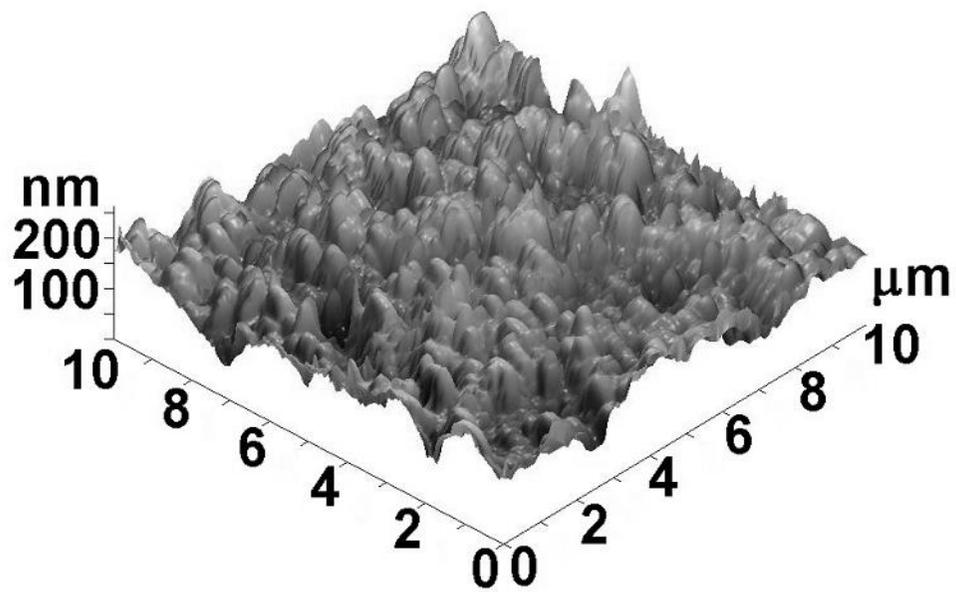

Fig.2a

Gregor et. al

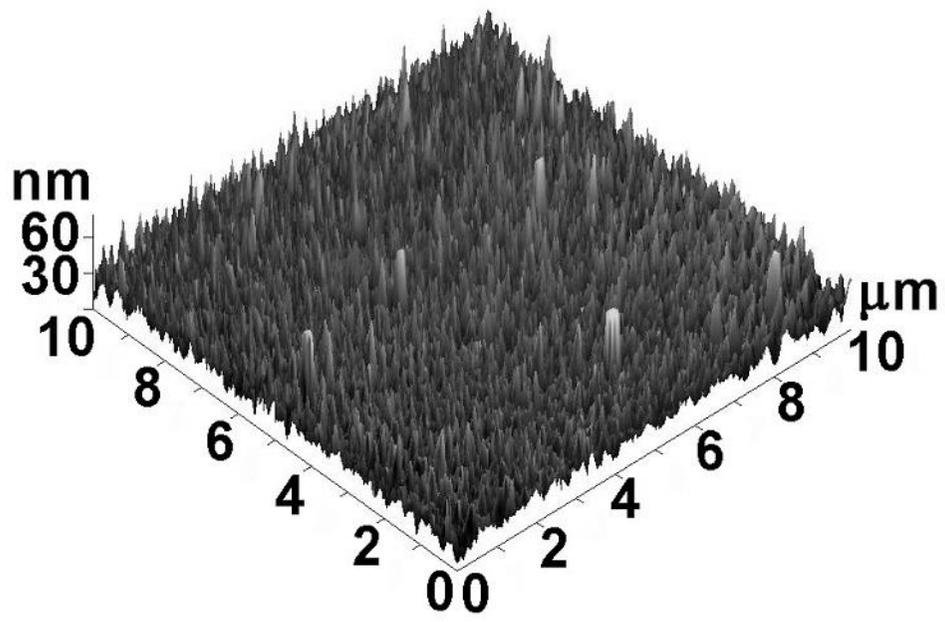

Fig.2b

Gregor et. al

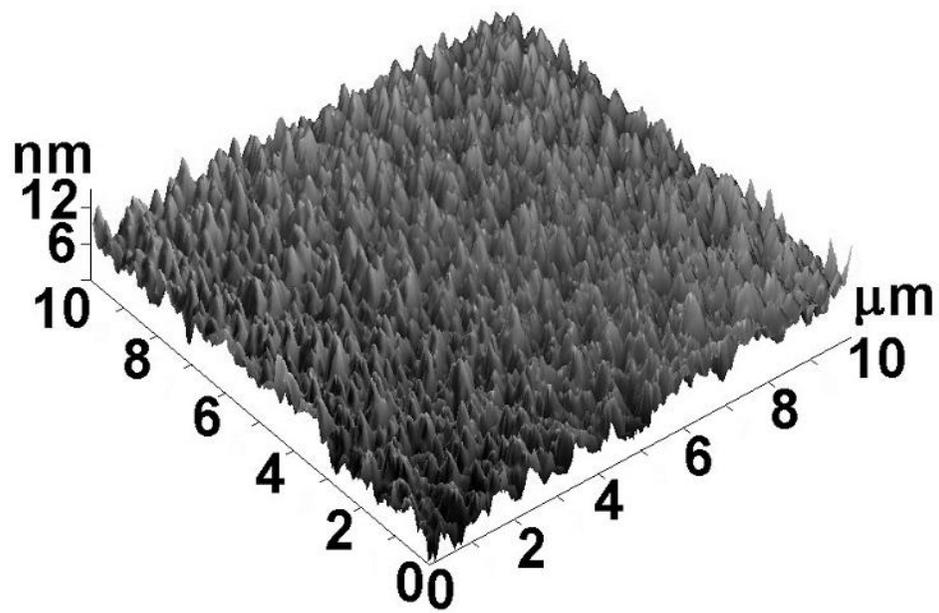

Fig.2c

Gregor et. al

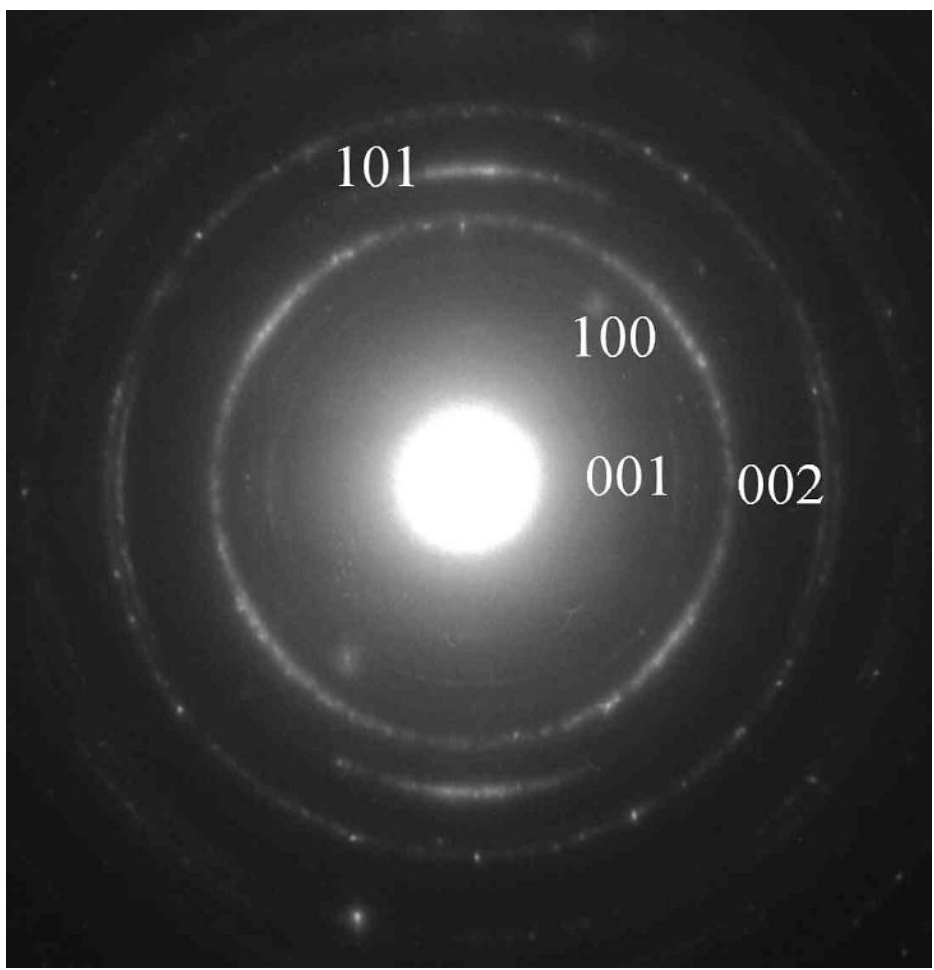

Fig.3

Gregor et. al

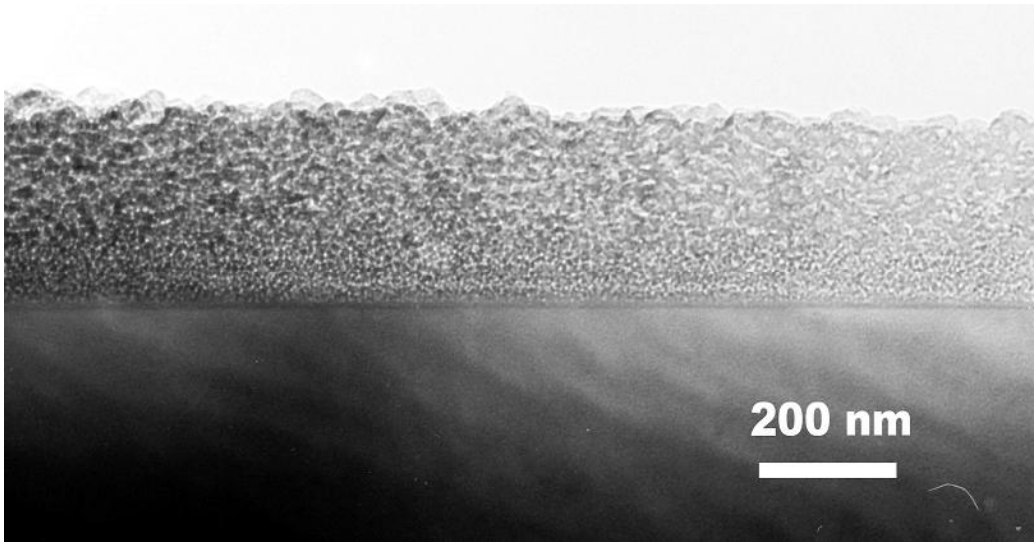

Fig.4a

Gregor et. al

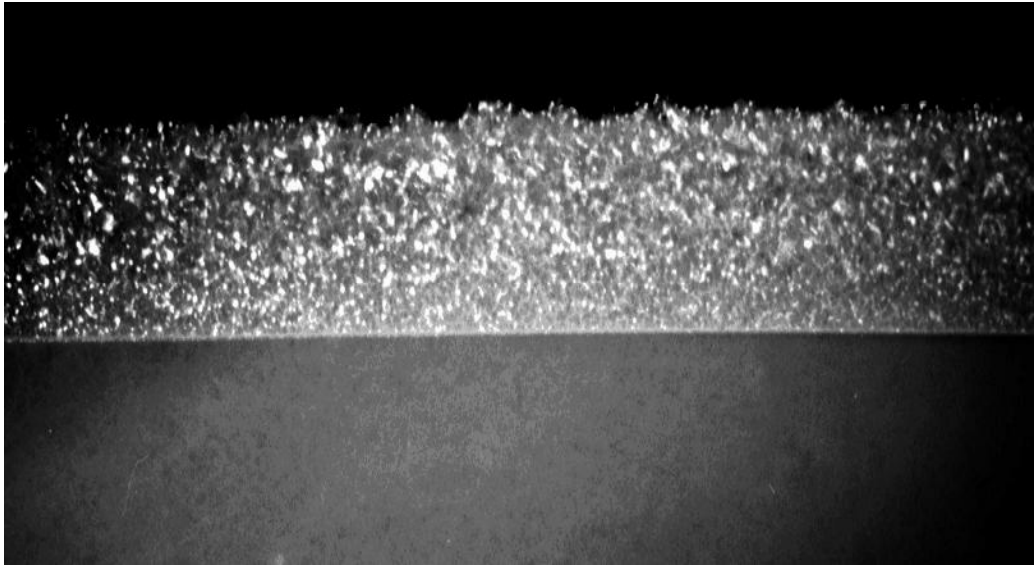

Fig.4b

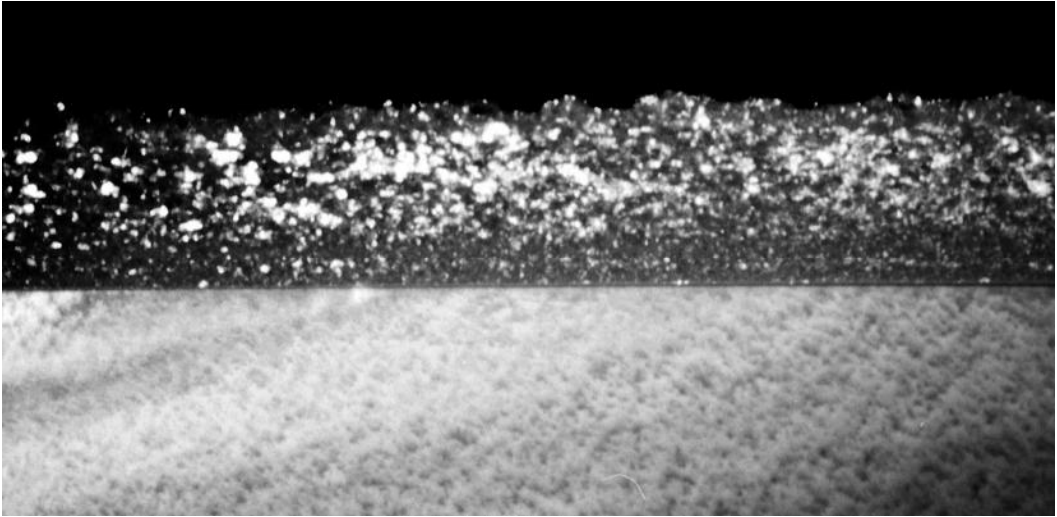

Fig.4c

Gregor et. al

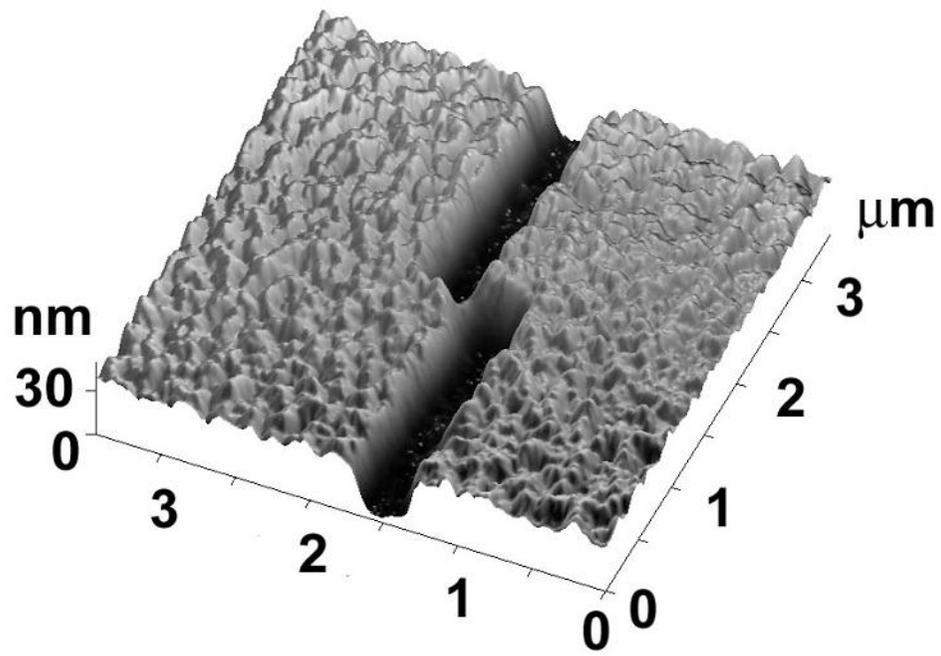

Fig.5

Gregor et. al

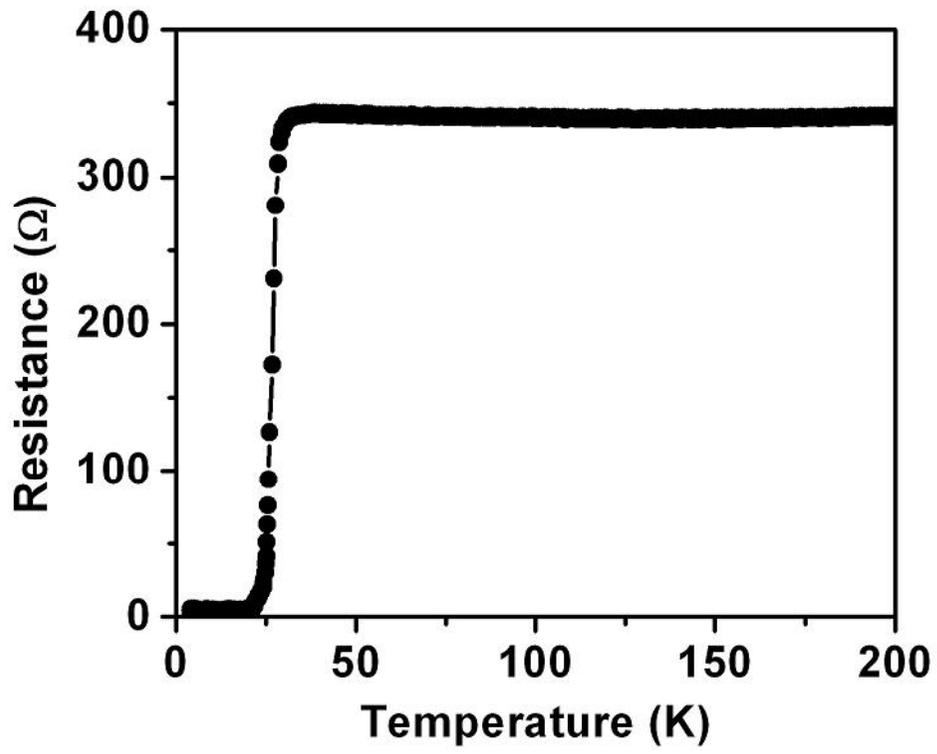

Fig.6a

Gregor et. al

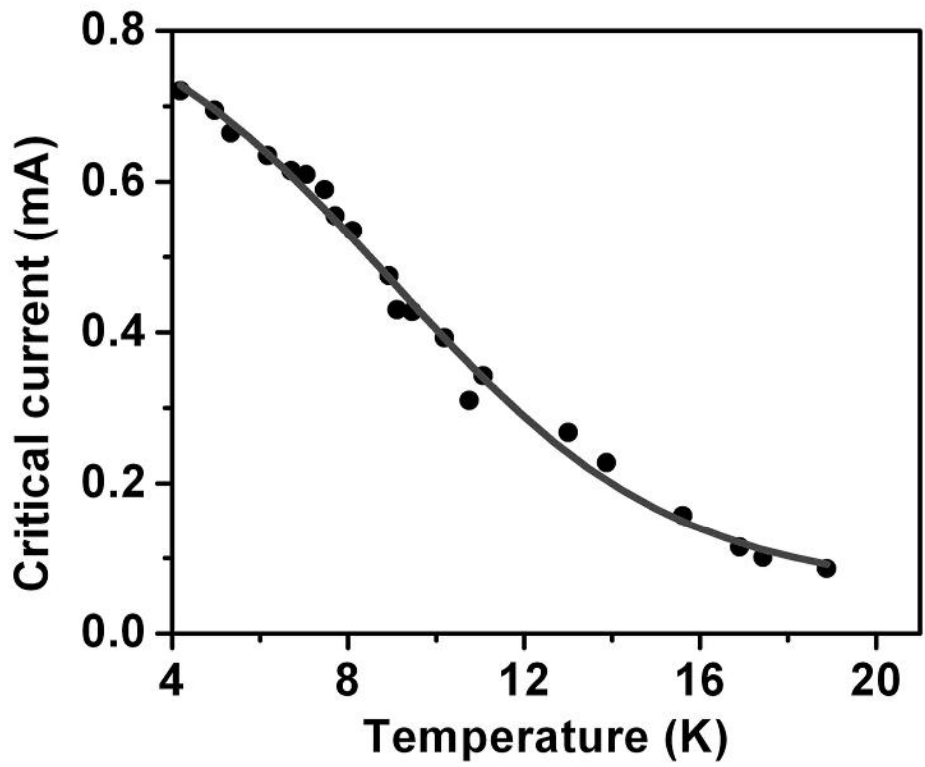

Fig.6b

Gregor et. al

| Sample | Ar atmosphere | Vacuum | Vaccum +Ar etch |
|---|---|---|---|
| Peak-to-peak, $R_{max}$ | 260.4 nm | 81.9 nm | 15.1 nm |
| RMS roughness, $R_q$ | 29.3 nm | 8.0 nm | 1.5 nm |

Table1

Gregor et. al